\documentclass[aps,pre,onecolumn,preprintnumbers,amsmath,amssymb,nofootinbib]{revtex4-2}

\usepackage{epsfig}
\usepackage{epstopdf}
\usepackage{booktabs}
\usepackage{lipsum}
\usepackage{multirow}
\usepackage{boldline}
\usepackage{xcolor}

\makeatletter
\renewcommand\section{\@startsection{section}{1}{\z@}
  {0.4cm \@plus1ex \@minus .2ex}{0.5cm}{\normalfont\small\bfseries\raggedright}}
\renewcommand\subsection{\@startsection{subsection}{2}{\z@}
  {.4cm \@plus1ex \@minus .2ex}{.5cm}{\normalfont\small\bfseries\raggedright}}
\renewcommand\subsubsection{\@startsection{subsubsection}{3}{\z@}
  {.4cm \@plus1ex \@minus .2ex}{.5cm}{\normalfont\small\itshape\raggedright}}
\makeatother

\usepackage{fancyhdr}
\fancypagestyle{custom}{
  \fancyhf{}
  \fancyhead[L]{\footnotesize\itshape K.\,P. Orzechowski, J. Sienkiewicz, P. Fronczak}
  \fancyhead[R]{\footnotesize\itshape Beyond global density in a stochastic sandpile}
  \fancyfoot[C]{\footnotesize\thepage}

}

\begin{document}
	\title{Beyond global density in a stochastic sandpile: the role of spatial signatures in the predictability of extremes}

\author{Kamil P. Orzechowski$^{1,*}$, Julian Sienkiewicz$^{1,2}$, and Piotr Fronczak$^{1}$}
\affiliation{$^{1}$Warsaw University of Technology, Faculty of Physics, Koszykowa 75, 00-662 Warsaw, Poland\\
$^{2}$Warsaw University of Technology, Centre for Credible AI, Rektorska 4, 00-614 Warsaw, Poland\\
$^{*}$Corresponding author: kamil.orzechowski2.dokt@pw.edu.pl}
\date{November 07, 2026}
	
\begin{abstract}
In self-organized critical systems the global load is a strong precursor of extreme events. In the two-dimensional Manna sandpile, grain density alone predicts extreme avalanches with an AUC of about 0.96. We ask whether density fully determines the system’s readiness for an extreme event, that is, whether microstates with the same number of grains are equivalent. Because density is such a strong predictor, a method that sees the full microstate can score well just by counting grains. We therefore compare only states with an identical number of grains. Even then, both a parameter-free statistic and a convolutional network distinguish states that precede an extreme avalanche from typical ones, while shuffled controls do not. The signal is a load contrast between the center and the dissipative boundaries, carried by the two lowest symmetric cosine modes. Their normalized energy is a single variable that needs no training. Together with density it recovers most of the gain that a network trained on the full microstate achieves over density alone. Readiness for an extreme avalanche therefore depends both on the total load and on how that load is distributed relative to the boundaries.
\end{abstract}
\maketitle

\section{Introduction}\label{sec:intro}

\noindent Self-organized criticality (SOC) describes slowly driven systems that reach a critical
state without any tuning of parameters and relax through avalanches with a power-law size
distribution~\cite{BTW1987,Jensen1998,Pruessner2012}. The concept has been applied to
earthquakes~\cite{BakTang1989}, solar flares~\cite{LuHamilton1991} and neuronal
activity~\cite{BeggsPlenz2003}, and in each of these settings the impact is dominated by extreme
events. The absence of a characteristic scale suggested that a large event has no
precursor~\cite{BakPaczuski1995,Geller1997}. Between large events, however, an SOC system is slowly
reloaded, so the state just before an extreme event may differ from a typical
one~\cite{Shapoval2022,Sapozhnikov2024}. Precursors of extreme avalanches in sandpile models have
been sought in time series of avalanche sizes and
activity~\cite{PepkeCarlson1994,Garber2009,Deluca2015,Sapozhnikov2024} and in features of the system
state~\cite{ShapovalShnirman2004,ShapovalShnirman2006,Shapoval2022}. In an experiment on a pile of
beads they were also sought in changes of its internal structure~\cite{Ramos2009}.

Here we study the Manna model~\cite{Manna1991} in its Abelian variant~\cite{Dhar1999}, which is
regarded as the representative of a universality class distinct from that of the deterministic model
of Bak, Tang and Wiesenfeld (BTW)~\cite{BenHur1996,Lubeck2004,Pruessner2012}. For this model the
question of a precursor has a partial answer. The mean height $q_t$, which is the global load of the
system, predicts extreme avalanches very well~\cite{Shapoval2022}. This fits the picture of SOC as
an absorbing-state phase transition in which density is the control
parameter~\cite{Dickman1998,Dickman2000}. Following Ref.~\cite{Shapoval2022}, we call a state with
elevated $q_t$ \emph{supercritical}. If $q_t$ told the whole story, microstates with the same number
of grains would be equally ready for an extreme event. Earlier studies indicated that they are not.
Before the largest avalanches the grains were seen to gather into clusters, and a precursor built
from these clusters could forecast large
avalanches~\cite{ShapovalShnirman2004,ShapovalShnirman2006,Shapoval2022}. It remained open whether
this component carries information independent of density or merely reflects
it~\cite{ShapovalShnirman2006}. To our knowledge it was also not known what form this information
takes or whether it can be expressed without tuned parameters.

The question is whether the number of grains alone is sufficient. Our focus is not on improving
the forecast, but on identifying the macroscopic quantities that describe the system's readiness for
an extreme event. A direct comparison of spatial arrangement is difficult because density is such a
strong predictor that a method using the full microstate, including a neural network, can perform well
simply by counting grains. This creates the possibility of shortcut learning~\cite{Geirhos2020} and
of attributing the prediction to the wrong physical mechanism~\cite{Carleo2019}. We therefore remove
density by thinning each configuration at random to a common number of grains and use shuffled
configurations to define the null level. We then examine the remaining spatial information with a
model-free statistic and with a convolutional neural network. The relevant scale is tested by
destroying structure below a prescribed length and by comparing different system sizes. Finally, we
expand the microstate in cosine modes, for which density is exactly the zero mode, and use this
representation to identify a small set of macroscopic variables. The network is used only as a
reference. Related work has already applied neural networks to avalanche prediction from sandpile
configurations~\cite{AlvezCanepa2026}.

\section{Methods}\label{sec:methods}

\subsection{Model and prediction task}\label{sec:model}

\noindent The Manna model is a stochastic sandpile on a square lattice of side $L$. At each time
step $t$ one grain is added to a randomly chosen site. A site that holds at least two grains becomes
unstable and passes two of them on, each to an independently drawn neighbor (von Neumann
neighborhood). Toppling continues as long as any site is unstable. Grains that topple across the
edge of the lattice leave the system, so the boundary is dissipative. Between avalanches every site
is either empty or occupied by a single grain, $h_t(x,y)\in\{0,1\}$ for $x,y\in\{0,\dots,L-1\}$. The
microstate $\mathbf{h}_t$ is thus a binary $L\times L$ image, and its global load $q_t$ is the mean
height, equal to the number of grains $k_t$ divided by the number of sites,
\begin{equation}\label{eq:q}
q_t=\frac{k_t}{L^{2}}=\frac{1}{L^{2}}\sum_{x,y}h_t(x,y).
\end{equation}
We keep the index $t$ for quantities followed along the run, such as $q_t$, and drop it when we
refer to a single, fixed image.

The avalanche size $s_t$ is the number of topplings triggered by the grain added at step $t$ to the
microstate $\mathbf{h}_t$. We call an avalanche extreme when $s_t>s^{*}L^{D}$, where $D=2.75$ is the
scaling exponent of avalanche sizes~\cite{Shapoval2022,Sapozhnikov2024} and $s^{*}=0.353$ is the
threshold for which Ref.~\cite{Shapoval2022} reports the most efficient prediction at $L=128$. We
choose this threshold so that the prediction from density alone can be compared directly with the
results of that work. We use it at every system size. A microstate is labeled $Y_t=1$ if the
avalanche that follows it is extreme and $Y_t=0$ otherwise. The label thus refers to an avalanche
that has not happened yet, and no image shows the avalanche it is labeled with. The threshold scales
with the system size in the same way as the cutoff of the avalanche distribution, so at every $L$ we
ask about roughly the same kind of event. Still, the frequency of extremes falls with $L$ faster
than scaling predicts. An avalanche distribution with exponent $\beta=14/11$~\cite{Shapoval2022}
implies that doubling $L$ reduces the number of extremes in a run of the same length to
$2^{-D(\beta-1)}\approx0.59$ of its previous value. In our data it falls to 0.33 between $L=32$ and
64 and to 0.49 between $L=64$ and 128 (Table~\ref{tbl1}, positives from both simulation phases
combined). This means that the smallest system still has clear finite-size corrections. Extremes are
relatively more frequent there, and a comparison with the larger systems does not concern exactly
the same kind of event.

\subsection{Data and evaluation}\label{sec:data_and_metrics}

\begin{figure}[h!]
  \centering
  \includegraphics[width=\linewidth]{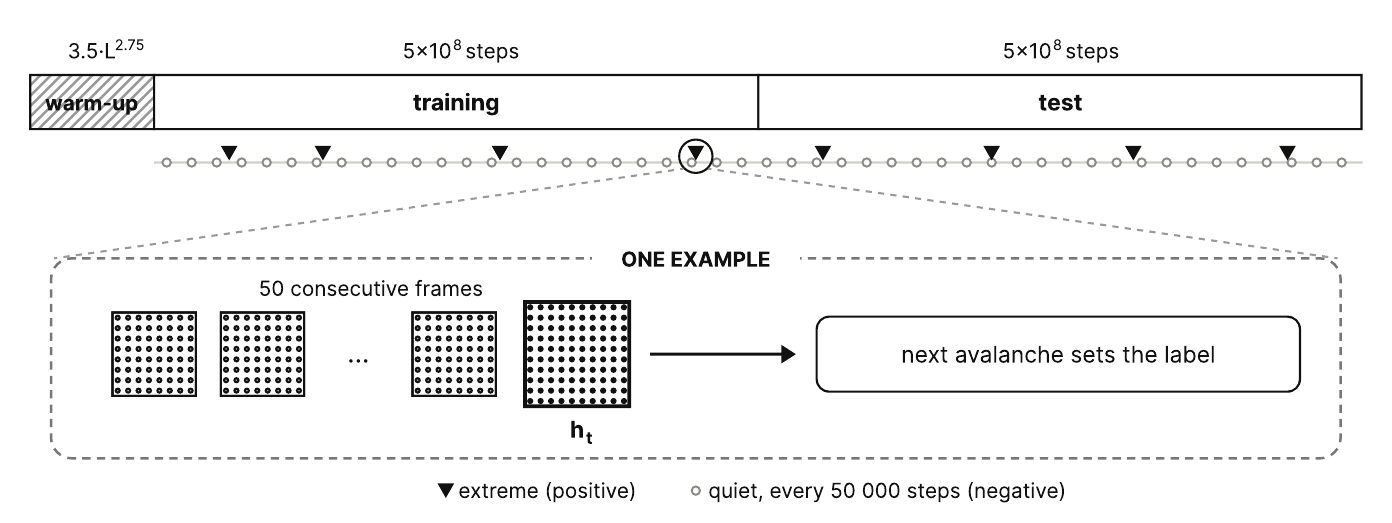}
  \caption{Data generation for each system size $L$. After a warm-up of $3.5\cdot L^{2.75}$ steps, one run continues through a training and a test phase of $5\times10^{8}$ steps each. Each example stores 50 frames ending in $\mathbf{h}_t$. Predictors use only $\mathbf{h}_t$, and the full window only sets the thinning level. All choices come from the training set, with a validation split stratified by label and drawn per seed (5 times).}\label{fig0}
\end{figure}

\noindent For each system size we warm up the simulation for $3.5\cdot L^{2.75}$ steps and then run
two phases of $5\times10^{8}$ steps each, a training phase and a test phase that continues the same
run. We call the sequence of all consecutive states of a phase a stream. The positives are all the
extremes that occurred. The negatives are quiet states taken every $50\,000$ steps, with no
condition on density or on history (Table~\ref{tbl1}). For every example we store a window of fifty
consecutive configurations, the last of which is the microstate $\mathbf{h}_t$ that precedes the
labeled avalanche. A single configuration is called a frame or an image. All predictors use only the
last frame, because the dynamics is Markovian and the earlier frames carry no information about the
avalanche size beyond what the last one contains. The data generation and collection process is presented on Fig. \ref{fig0}. The full windows are used only to set the thinning
level (Section~\ref{sec:thinning}). The test set is used only for the final evaluation. Every
constant and every choice made later in the paper comes from the training set or from a validation
part split off from it (20\%, stratified by label and drawn separately for each of the five random
seeds). 

\begin{table}[h!]
\begin{tabular*}{\linewidth}{@{\extracolsep{\fill}}llllll@{}}
\toprule
$L$ & training: total & training: positives & test: total & test: positives & negatives per phase \\
\midrule
32  & 18\,114 & 8\,114 & 18\,108 & 8\,108 & 10\,000 \\
64  & 12\,674 & 2\,674 & 12\,741 & 2\,741 & 10\,000 \\
128 & 11\,339 & 1\,339 & 11\,309 & 1\,309 & 10\,000 \\
\bottomrule
\end{tabular*}
\caption{Collected examples for each system size $L \in \{32, 64, 128\}$.\label{tbl1}}
\end{table}
\noindent \textbf{AUC}: For a predictor f mapping each state x to a scalar score f(x),
\begin{equation}
\mathrm{AUC}=\Pr\!\left(r^{+}>r^{-}\right)+\tfrac12\Pr\!\left(r^{+}=r^{-}\right),
\label{eq:auc}
\end{equation}
where $r^{+} = f(x^{+})$ and $r^{-} = f(x^{-})$ are the scores of a randomly drawn positive state $x^{+}$ and an independently drawn negative state $x^{-}$~\cite{Fawcett2006, HanleyMcNeil1982}. This quantity does not depend on the class proportions, and a score
that returns the same number for every state produces only ties, which gives $\mathrm{AUC}=1/2$.

\noindent \textbf{Prediction error}: To compare with earlier
work~\cite{Shapoval2022,Sapozhnikov2024} we also use the error
$\varepsilon=\nu+\tau$~\cite{Molchan1997,KeilisBorok2003}, the sum of the fraction of missed
extremes $\nu$ and the false-alarm rate $\tau$, i.e., the fraction of quiet states on which an alarm is raised, at a single threshold. The threshold is
chosen by minimizing $\varepsilon$ on the training set and is then applied to the test set. Computed
on the full test stream, this error is directly comparable with the results of
Ref.~\cite{Shapoval2022}. Computed on the sets of Table~\ref{tbl1}, it is used to compare
predictors. Neither metric depends on the class proportions, which is why we keep all of the rare
positives and subsample the abundant negatives.

\subsection{Removing density: thinning and conditioning on the number of grains}\label{sec:thinning}

\noindent Because $q$ is a good predictor, no statement about space that rests on raw images can
separate arrangement from density. We therefore remove density from the data.

\noindent \textbf{Thinning}: From every image we remove grains at random, without looking at the
label, until exactly $K$ remain. Parameter $K$ is a single number per system size, shared by both classes, all
examples and all predictors. It equals the smallest number of grains found in any frame of the
training set. This level is not tuned. It is forced on us, because grains may be removed but must
not be added. Adding a grain would mean inventing its position, which is exactly the quantity we are
asking about. From the last frames, on which the predictors operate, thinning removes on average
between about 1\% and 8\% of the grains, depending on the system size (Table~\ref{tbl2}). Test
examples whose last frame has fewer than $K$ grains are discarded. After thinning every image has
density $K/L^{2}$, the same in both classes, so a predictor that uses density alone returns a
constant and, by the tie term of Eq.~(\ref{eq:auc}), has an AUC of exactly $1/2$. A binary
image with a fixed number of ones remains, and two such images can differ only in where the ones
are. During training the thinning is drawn anew every time an example is read. During evaluation
each example is always thinned in the same way, which means that all models are evaluated on
identical images.

\begin{table}[h!]
\setlength{\tabcolsep}{3pt}
\begin{tabular*}{\linewidth}{@{\extracolsep{\fill}}llllll@{}}
\toprule
$L$ & frames & $K$ & removed: worst frame & removed: mean & discarded test examples \\
\midrule
32 & 905\,700 & 659     & 14.4\% & 7.2\% & 0 of 18\,108 \\
64 & 633\,700 & 2\,790  & 6.1\%  & 2.9\% & 0 of 12\,741 \\
128 & 566\,950 & 11\,464 & 2.6\%  & 1.1\% & 1 of 11\,309 \\
\bottomrule
\end{tabular*}
\caption{The forced thinning level $K$ and the fraction of grains that thinning removes, computed
over all frames of all training windows. In the last frames, on which the predictors operate, this
fraction is slightly larger. The last column gives the number of test examples discarded because
their last frame has fewer than $K$ grains.}\label{tbl2}
\end{table}

\noindent \textbf{Null level}: We repeat every run on shuffled images, in which the grains are
first scattered uniformly over the lattice and only then thinned to the same $K$. The architecture,
the training schedule and the seeds of the random number generator stay the same. The result of such
a run is the null level, the amount of signal that the training and evaluation procedure can produce
by itself.

\noindent \textbf{Conditioning on the number of grains}: Thinning equalizes the number of grains,
but it removes more grains from a frame that had more of them. This leaves a weak trace of the
original number of grains $k$, and the null level does not capture this trace. For this
reason we also compare only states with an identical original number of grains $k$. Within such a
group exactly the same number of grains is removed from every image, so the procedure is
statistically identical for both classes and cannot separate them by itself. Also, $q$ is constant
within a group, so its conditional AUC is exactly $1/2$. We report the conditional AUC computed
jointly over all pairs of a positive and a negative with the same number of grains,
\begin{equation}
\mathrm{AUC}_{|k}=\frac{\sum_{k}\sum_{i\in P_k}\sum_{j\in N_k}
\left[\mathbf{1}(r_i>r_j)+\tfrac12\,\mathbf{1}(r_i=r_j)\right]}{\sum_{k}|P_k|\,|N_k|},
\label{eq:auck}
\end{equation}
where $P_k$ and $N_k$ are the positives and the negatives with $k$ grains, $r_i$ is the score of example $i$, and $\mathbf{1}(\cdot)$ equals 1 when the condition holds and 0 otherwise. This control does not
require thinning, so we also apply it to raw images.

\subsection{Spatial observables}\label{sec:observables}
\noindent After the number of grains is fixed, the remaining information in an image is the spatial
arrangement of the grains. We examine this arrangement in two ways. The first is a quantity computed
directly from the image, without machine learning. The second progressively removes structure below a
given scale and tests which scales contribute to the network prediction.

\subsubsection{Excess block-density variance}
\noindent We divide the lattice into $B=(L/\ell)^{2}$ non-overlapping blocks of side $\ell$ and
compute the density of each, $\rho_b=\frac{1}{\ell^{2}}\sum_{(x,y)\in b}h(x,y)$. The variance of
these densities measures how uneven the arrangement is at scale $\ell$~\cite{Torquato2003},
\begin{equation}
V_\ell=\frac{1}{B}\sum_{b=1}^{B}\left(\rho_b-q\right)^{2},
\label{eq:Vl}
\end{equation}
where $q$ is the density of the image on which $V_\ell$ is computed. By varying $\ell$ from $L/2$ to
$L/32$ we probe the structure at successive scales. The blocks must tile the lattice exactly, so the
scan has a resolution of one octave. The raw difference in $V_\ell$ between the classes cannot be
compared across scales or across system sizes, because the block-density variance depends on the
block size even when the grains are placed completely at random. This is why we express the
difference between the classes in units of the variance that a random placement of the same number
of grains would give at that scale,
\begin{equation}\label{eq:delta}
\Delta(\ell/L)=\frac{\langle V_{\ell}\rangle_{+}-\langle V_{\ell}\rangle_{-}}{E_{\text{rand}}},\qquad
E_{\text{rand}}=\frac{q(1-q)\left(L^{2}-\ell^{2}\right)}{\ell^{2}\left(L^{2}-1\right)} ,
\end{equation}
where $\langle V_{\ell}\rangle_{+}$ and $\langle V_{\ell}\rangle_{-}$ are the means over positives
and over negatives, and $E_{\text{rand}}$ is the expected value of $V_\ell$ for the same number of
grains scattered over the lattice completely at random (derived in Appendix~\ref{app:scans}). The
random level is used here as a unit and is not meant as a model of the system. In models of the
Manna class the density fluctuations of critical states are suppressed relative to a random
arrangement~\cite{Hexner2015}. The scan is computed on thinned images, so $q=K/L^{2}$ and a single
$E_{\text{rand}}$ applies to both classes. Thinning weakens the excess by an amount that depends on
the system size but not on the scale. For this reason we divide each curve by its own maximum and
compare, between sizes, the \emph{shape} of $\Delta(\ell/L)$ and the position of its maximum.

\subsubsection{Destruction scan}
\noindent The block-density variance tells us at which scale the classes differ in the data, but not
which scale the network uses. To put the same question to the network on the same axis, we divide
the lattice of each thinned test image into blocks of side $\ell$ and scatter the grains of every
block at random over the sites of that block. This keeps the number of grains in each block, and
with it all structure at scales larger than $\ell$, and removes only the arrangement of the grains
inside the blocks. It is a counterpart of the surrogate-data method~\cite{Theiler1992}. The network
trained on thinned images is evaluated on the modified images without retraining, and we measure the
fraction of its advantage over the random level that is lost,
\begin{equation}\label{eq:lost}
F(\ell/L)=\frac{\mathrm{AUC}_{0}-\mathrm{AUC}_{\ell}}{\mathrm{AUC}_{0}-\tfrac12},
\end{equation}
where $\mathrm{AUC}_{0}$ is the result on intact images and $\mathrm{AUC}_{\ell}$ the result on
images destroyed below scale $\ell$. The curve $F$ is cumulative, so the scale of the signal shows
up in its increments. The difference $F(2\ell/L)-F(\ell/L)$ is the fraction of the network's
advantage that comes from structure with sizes between $\ell$ and $2\ell$. The normalization relates
the loss to the advantage of the given model, which makes different sizes comparable. The receptive
field of the network grows in proportion to $L$ (Section~\ref{sec:predictors}), so the scan shows
which scales are used by a network that was allowed to use them. As evidence about scale, the scan
is not independent of this design choice.

\subsection{Modal decomposition}\label{sec:modes}
\noindent The tools of the previous section show at which scale the spatial information lies, but
they do not say what shape it has, and they rely on thinning. The modal decomposition fills both
gaps. We write the microstate in a basis in which density is one coordinate and each of the others
describes a spatial pattern with a definite scale and orientation.

Let $\mathbf{h}$ be the microstate stored in the last frame of a window, without thinning, and
$h(x,y)$ its value at site $(x,y)$. We expand the microstate in an orthonormal cosine basis (the
two-dimensional DCT-II)~\cite{Strang1999},
\begin{equation}\label{eq:dct}
h(x,y)=\sum_{m,n=0}^{L-1}a_{mn}\phi_{mn}(x,y),\qquad \phi_{mn}(x,y)=\phi_m(x)\phi_n(y),
\end{equation}
with amplitudes $a_{mn}=\sum_{x,y}h(x,y)\,\phi_{mn}(x,y)$, where
\begin{equation}\label{eq:basis}
\phi_m(x)=c_m\cos\!\left[\frac{\pi m}{L}\left(x+\tfrac12\right)\right],\qquad
c_0=\frac{1}{\sqrt{L}},\quad c_m=\sqrt{\frac{2}{L}}\ (m>0),
\end{equation}
and the indices $m,n=0,\dots,L-1$ label the modes. The function $\phi_m(x)$ changes sign $m$ times
across the system, so small indices correspond to long-wavelength modes and large indices to
fine-scale structure. We choose this basis because its zero mode \emph{is} the density, $a_{00}=Lq$,
while every mode $(m,n)\neq(0,0)$ describes only the arrangement of the grains. This separation is
algebraic and not statistical, which is why we also assess the predictive power of the modes within
groups of states with an identical number of grains (Section~\ref{sec:thinning}). We also do not
claim that this basis diagonalizes the dynamics. It diagonalizes the lattice Laplacian with a
reflecting boundary~\cite{Strang1999}, whereas the boundary of the model is dissipative. The
contribution of a pattern to the microstate is measured by the mode energy $a_{mn}^{2}$. The total
energy of the nonzero modes equals $L^{2}q(1-q)$, which follows from Parseval's theorem and from
$h^{2}=h$. This total is fixed by $q$ alone and carries no new information. Only the way this energy
\emph{is distributed} among the modes is new, so we work with the normalized mode energies
\begin{equation}\label{eq:Rmn}
R_{mn}=\frac{a_{mn}^{2}}{L^{2}q(1-q)},\qquad \sum_{(m,n)\neq(0,0)}R_{mn}=1 ,
\end{equation}
that is, with the share of a given mode in the total inhomogeneity of the microstate. Each mode is
treated separately as a one-dimensional predictor and assessed by $\mathrm{AUC}(R_{mn})$. The mode
with the largest AUC is selected on the training set only, and the degenerate modes $(m,n)$ and
$(n,m)$, which map onto each other under a rotation by $90^{\circ}$, are summed. The resulting
quantity is invariant under rotations and reflections of the lattice, the same symmetries that the
dynamics of the model has.

\subsection{Predictors}\label{sec:predictors}

\noindent The block-density variance $V_\ell$ and the normalized mode energies $R_{mn}$ are used
directly as scores and have no parameters fitted to the data. The scale $\ell$ and the pair of modes
are chosen on the training set. We combine density and the selected mode energy by logistic
regression without regularization, on features standardized with the statistics of the training set,
\begin{equation}\label{eq:logit}
\operatorname{logit}\Pr(Y=1)=\theta_0+\theta_q\tilde q+\theta_R\tilde R_2 ,
\end{equation}
where the logit is the logarithm of the odds $\Pr(Y=1)/\Pr(Y=0)$, $R_2$ is the sum of the normalized energies of the degenerate pair picked by the selection
rule of Section~\ref{sec:modes}, and the tilde denotes standardization. The fitted coefficients $\theta_q$ and $\theta_R$ give the change in the log-odds of an extreme per one standard deviation of $q$ and of $R_2$, and $\theta_0$ is the intercept. The model is fitted on the
training set and applied unchanged to the test set.

The reference is a convolutional neural network~\cite{LeCun2015}. It consists of convolutional
blocks with stride 2 (convolution, batch normalization~\cite{Ioffe2015}, ReLU), a head that computes
the mean and the standard deviation over positions for every channel of the last feature map, and a
linear layer. The number of blocks grows with the system size (3, 4, 5 for $L=32$, 64, 128), so that
the receptive field stays a fixed fraction of the system side instead of a fixed number of sites.
The head averages over positions, so by construction the network is only weakly sensitive to
\emph{where} in the system a given structure lies. We regard it as a typical general-purpose model.
It gives a lower estimate of what can be extracted from the microstate, not an upper bound. We train
five models from independent initializations and average their \emph{metrics}, never their scores.
Training details and learning curves are given in Appendix~\ref{app:net}.

\subsection{Statistics}\label{sec:stats}
\noindent Confidence intervals come from a bootstrap~\cite{Efron1993} over test examples (95\%
percentile intervals from 2000 resamples, paired when predictors are compared). In comparisons with
the network the bootstrap is run with the five training seeds held fixed, so it does not include the
spread between seeds. We report that spread separately as a standard deviation ($\pm$). The $p$
values come from a one-sided permutation test with 2000 label permutations, for which the smallest
attainable value is $0.0005$~\cite{Phipson2010}. The bootstrap assumes that the examples are
independent. For the positives this is supported by the coefficient of variation of the intervals
between extremes, 0.98-1.05, which is consistent with a Poisson process. For the negatives the
autocorrelation of $q_t$ at the sampling interval ($50\,000$ steps) is indistinguishable from zero
(its absolute value does not exceed 0.02 at any size).

\section{Results}\label{sec:results}

\noindent We first establish density as the baseline predictor. We then characterize the spatial
signal by its scale and shape and reduce the dominant pattern to the single variable $R_2$. The final
comparison asks how well the pair $(q,R_2)$ reproduces the performance of a network trained on the full
microstate. Unless stated otherwise, the reported values come from the test set.

\subsection{Baseline: density as a predictor of extremes}\label{sec:baseline}

\begin{figure}[h!]
  \centering
  \includegraphics[width=\linewidth]{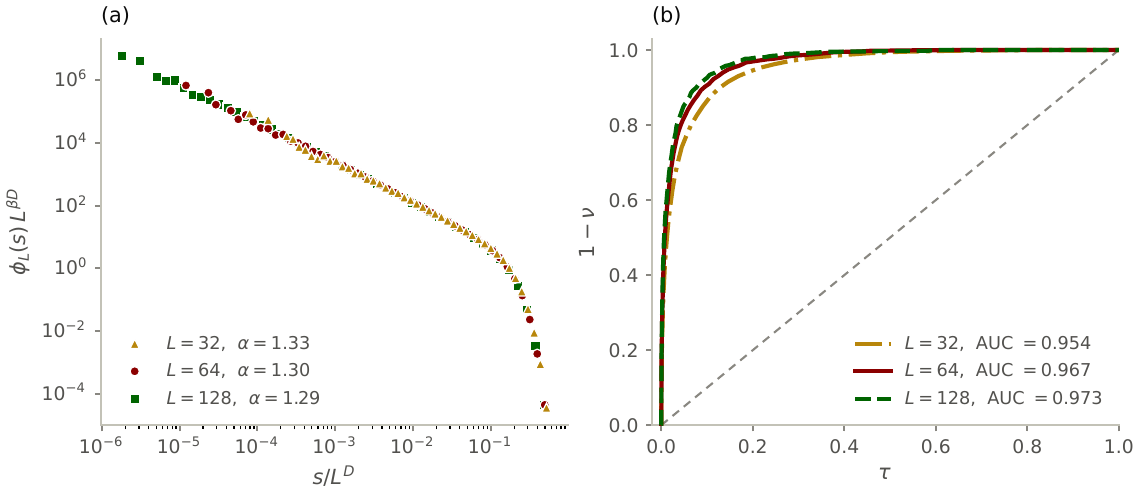}
  \caption{(a) Probability densities of avalanche sizes $\phi_L(s)$ after both axes are rescaled
  with the system size, using $D=2.75$ and $\beta=14/11$~\cite{Shapoval2022}. The fitted power-law
  exponents $\alpha$ are given in the legend. (b) ROC curves for the density $q_t$ alone, which
  separates states before an extreme from quiet states with
$\mathrm{AUC}\approx0.95$-$0.97$. The figures are outcomes of simulation of the algorithm presented by Shapoval et al. \cite{Shapoval2022}, and prepared by the authors of this manuscript.}\label{fig1}
\end{figure}

\begin{table}[h!]
\begin{tabular*}{\linewidth}{@{\extracolsep{\fill}}lllllll@{}}
\toprule
$L$ & $\mathrm{AUC}(q_t)$ & $\alpha$ & $\delta_1$ & $q^{*}$ & $\varepsilon^{*}(q_t)$ &
$\varepsilon$ at $q^{*}$ \\
\midrule
32  & 0.9542 & 1.330 & 2.071 & 0.7002 & 0.228 & 0.228 \\
64  & 0.9674 & 1.298 & 2.190 & 0.7058 & 0.192 & 0.195 \\
128 & 0.9725 & 1.293 & 2.222 & 0.7099 & 0.172 & 0.174 \\
\bottomrule
\end{tabular*}
\caption{Baseline: prediction from density alone on the full test stream. The presented quantities are described as follows: (i) $\alpha$ is the exponent
of the avalanche-size distribution (linear regression on a log-log scale, 49 logarithmic bins, range
$10<s<0.1L^{D}$), (ii) $\delta_1$ is the absolute slope of the linear dependence of the error $\varepsilon$ on the
rescaled extreme threshold $s^{*}$~\cite{Shapoval2022} (linear regression for $0.05<s^{*}<0.3$, with
the density threshold chosen separately for each $s^{*}$), (iii) $q^{*}$ is the density threshold that
minimizes the error in the training phase (an alarm is raised when $q_t\ge q^{*}$), and (iv)
$\varepsilon^{*}(q_t)$ is the smallest error attainable by a threshold rule on density alone. Its
threshold is chosen on the test stream itself, so it is a lower bound and not a predictor. The last
column is the prediction error at the threshold $q^{*}$.}\label{tbl3}
\end{table}

\noindent We begin by checking that our simulations reproduce the known result that the density
$q_t$ alone predicts extreme avalanches well. The avalanche-size distributions for the three system
sizes collapse after rescaling (Fig.~\ref{fig1}a), as described for the Manna
model~\cite{Sapozhnikov2024}, and density separates states before an extreme from quiet states with
an AUC of $0.95$ to $0.97$ (Fig.~\ref{fig1}b, Table~\ref{tbl3}). Reference~\cite{Shapoval2022}
reports, for $L=128$ and the same threshold $s^{*}$, the values $\mathrm{AUC}=0.946$, $q^{*}=0.7098$
and $\varepsilon=0.205$, obtained from 131 test events.

Our results at $L=128$ ($\mathrm{AUC}=0.973$, $q^{*}=0.7099$, $\varepsilon=0.174$, 1309 events)
agree with these. The rate of extremes agrees as well (131 in
$5\cdot10^{7}$ steps against 1309 in $5\cdot10^{8}$)\footnote{Reference~\cite{Shapoval2022} gives
the length of the test set in two ways, $50\times10^{7}$ and $50\cdot10^{6}$ steps. We take the
value stated together with the number of events.}, and so does the slope of the linear decrease of
the error with the threshold $s^{*}$ ($\delta_1\approx2.23$ in Ref.~\cite{Shapoval2022} for larger
systems, while our values grow with $L$ up to 2.22, Table~\ref{tbl3}). This confirms that the
simulation and the data collection are correct, and it sets the baseline for the comparisons that
follow.

The strength of this predictor also makes the question about the arrangement harder to answer. A
convolutional network trained on the collected examples (Table~\ref{tbl1}) without thinning reaches
an AUC of $0.990$ to $0.996$, which is more than density alone. On raw images, however, this surplus
cannot be cleanly isolated, because by far the largest part of the score can be obtained just by
counting grains~\cite{Geirhos2020}. In the next step we therefore remove density from the data.

\subsection{Spatial information at a fixed number of grains}\label{sec:constant}

\noindent Thinning (Section~\ref{sec:thinning}) brings every image to the same number of grains, so
density is no longer a variable and its AUC is $1/2$ by construction. If the classes can still be
told apart, only the arrangement of the grains can be responsible. We test this with two tools, the block-density variance at $\ell/L=1/4$, computed directly from the image, and a convolutional network
(Table~\ref{tbl4}). We fix the scale before looking at the test set: $\ell/L=1/4$ is where the excess of Section~\ref{sec:scale} peaks on the training set.

\begin{table}[h!]
\setlength{\tabcolsep}{3pt}
\begin{tabular*}{\linewidth}{@{\extracolsep{\fill}}lllllll@{}}
\toprule
$L$ & $q$, raw & $V_{L/4}$, raw & $V_{L/4}$, thinned & $V_{L/4}$, null level &
network, thinned & network, null level \\
\midrule
32  & 0.954 & 0.691 & 0.680 & $0.500\pm0.003$ & $0.8042\pm0.0012$ & 0.4999 ($p=0.52$) \\
64  & 0.967 & 0.744 & 0.731 & $0.499\pm0.006$ & $0.8567\pm0.0024$ & 0.4960 ($p=0.93$) \\
128 & 0.974 & 0.782 & 0.779 & $0.498\pm0.011$ & $0.8658\pm0.0050$ & 0.5017 ($p=0.33$) \\
\bottomrule
\end{tabular*}
\caption{Spatial signal at a fixed number of grains (AUC on the sets of Table~\ref{tbl1}). The quantity $V_{L/4}$
is the block-density variance at $\ell/L=1/4$, and the null level is measured on shuffled images.
The spread $\pm$ refers to five random shuffles (block variance) or five training
seeds (network). The $p$ values come from the permutation test.}\label{tbl4}
\end{table}

Both tools distinguish the classes at a fixed number of grains, the block variance with an AUC of
$0.68$ to $0.78$ and the network with $0.80$ to $0.87$. The block variance loses almost nothing
through thinning, and the null level coincides with the random level in both cases. The signal is
therefore not produced by the training and evaluation procedure itself. Nor is it the trace of the
original density that thinning leaves behind. Within groups of states with an identical original
number of grains the AUC (Eq.~(\ref{eq:auck})) does not drop. It rises to $0.72$-$0.85$ for the block variance and to
$0.89$-$0.93$ for the network. The groups that contain both classes cover from 78\% (at $L=128$) to
95\% (at $L=32$) of the positives, so the conditioning covers most of the data.

At an identical number of grains, then, the arrangement alone separates states before an extreme
from quiet states. The number of grains does not fully describe the state. Much of the signal is
already carried by a statistic that involves no training, so the effect is not an artifact of a
flexible model.

\subsection{Scale of the signal: a fixed fraction of the system size}\label{sec:scale}

\begin{figure}[h!]
  \centering
  \includegraphics[width=\linewidth]{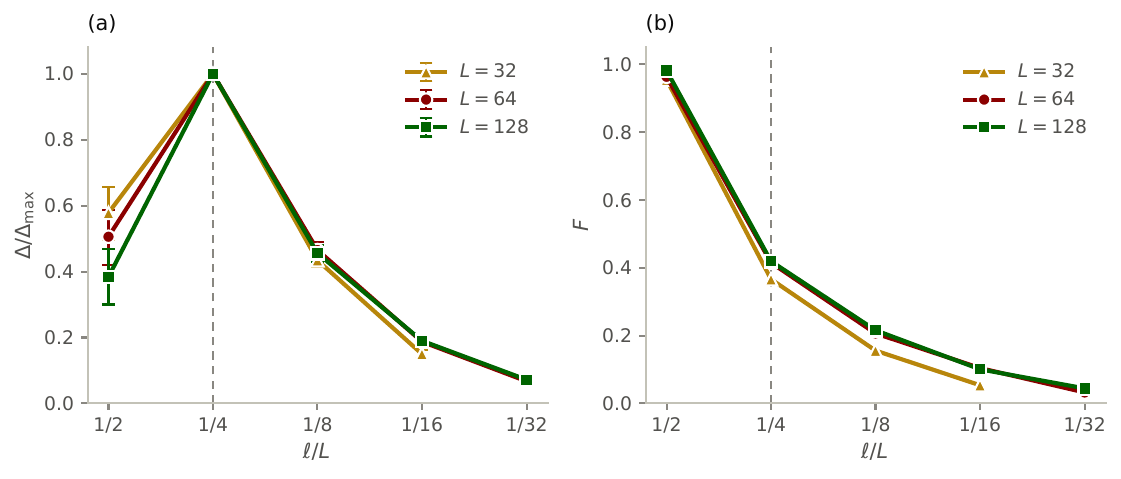}
  \caption{(a) Shape of the excess block variance, $\Delta(\ell/L)$ divided by its own maximum.
  Error bars are 95\% bootstrap intervals. At every size the maximum falls at $\ell/L=1/4$ (dashed
  line). (b) Lost fraction of the network's advantage $F(\ell/L)$ after the structure below scale
  $\ell$ is destroyed. The curve is cumulative, so the scale of the signal is given by its largest
  increment, between $\ell/L=1/4$ and $1/2$. In both panels the scale decreases to the right, from
  $\ell/L=1/2$ to $1/32$ (at $L=32$ to $1/16$).}\label{fig2}
\end{figure}

\noindent We now ask at what spatial scale the information that separates the classes is found. We use both tools of Section~\ref{sec:observables}, the excess block variance $\Delta(\ell/L)$, which
describes the data, and the destruction scan $F(\ell/L)$, which describes the network
(Fig.~\ref{fig2}).

The maximum of $\Delta$ falls at $\ell/L=1/4$ for all three sizes, both on the training set, from
which we read it off, and on the test set. This corresponds to blocks of $8\times8$ to $32\times32$
sites, a fixed fraction of the system size instead of a fixed number of sites. Near the maximum the
shapes of the curves agree. The largest discrepancy between sizes, at $\ell/L=1/2$, falls within the
confidence intervals when the scan is repeated at a fixed original number of grains, which marks it
as a trace of thinning and not a property of the arrangement (Appendix~\ref{app:scans}). The
position of the maximum is known only to within an octave, because the scan covers only scales that
differ by a factor of two.

The destruction scan leads to the same conclusion. Destroying the structure below $L/4$ takes away
less than half of the network's advantage. More than half (0.55 to 0.59) disappears only when the
destruction is extended to $L/2$, so it comes from structure with sizes between $L/4$ and $L/2$.

The model-free scan thus places the signal at a scale proportional to $L$, and the destruction scan
shows that a network whose receptive field scales with the system uses the same band. That the scale
of the precursor is proportional to $L$ had been conjectured earlier~\cite{ShapovalShnirman2006}.
The scan alone does not tell us whether this proportionality reflects the self-similarity of the
critical state or a load profile imposed by the dissipative boundary, because the distance from the
boundary also scales with $L$. A similar doubt was raised for the BTW model~\cite{Garber2009}. We
return to this point in the discussion.

\subsection{Shape of the signal: a contrast between the center and the boundaries}\label{sec:pattern}

\begin{figure}[h!]
  \centering
  \includegraphics[width=\linewidth]{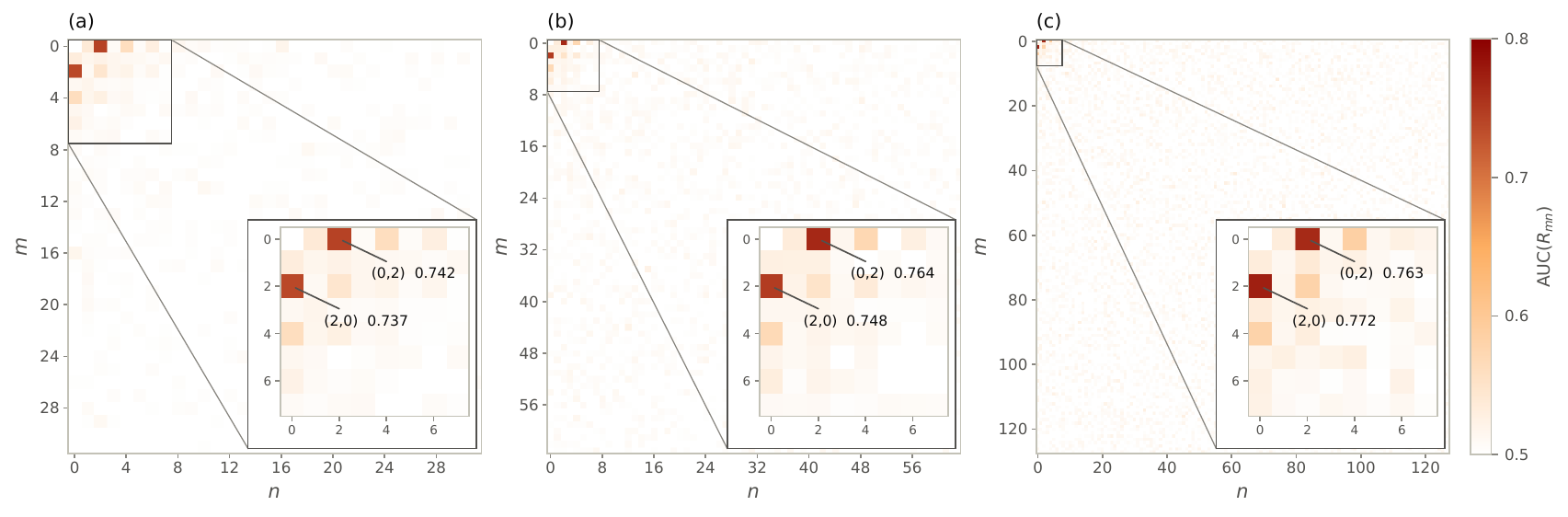}
  \caption{AUC of the normalized energy $R_{mn}$ of every mode on the training set for (a) $L=32$, (b) $L=64$ and (c) $L=128$. Each panel shows the full spectrum, and the inset enlarges the modes with $m,n<8$, the region framed in the upper left corner. The numbers next to the selected pair are its
  AUC values. White marks a mode that carries no information (AUC $=0.5$) and darker red a mode with higher energy before an extreme. No mode falls below AUC $=0.46$, so values under $0.5$ are also drawn in white. The mode $(0,0)$ corresponds to density and is not assessed here. At every $L$ the
  same pair, $(2,0)$ and $(0,2)$, stands out.}\label{fig3}
\end{figure}

\noindent The scale does not yet tell us what shape the structure behind the signal has. To find
out, we expand the microstates in cosine modes (Section~\ref{sec:modes}) and check for each mode how
well its normalized energy $R_{mn}$ distinguishes the classes (Fig.~\ref{fig3}).

At every size the two strongest modes are the same pair, $(2,0)$ and $(0,2)$, with an AUC of $0.74$
to $0.77$ on the training set, while no other mode exceeds $0.59$. The pair was identified at $L=32$
and then kept fixed. This means that at $L=64$ and $L=128$ the pair was not selected on the data on
which it is assessed, and the corresponding panels of Fig.~\ref{fig3} are a confirmation and not a
discovery.

The pattern has a simple geometric interpretation. The function
$\phi_2(x)\propto\cos\left[2\pi\left(x+\frac12\right)/L\right]$ is positive near both boundaries and
negative in the middle, and it changes sign at a distance $L/4$ from each boundary. The amplitude of
the mode $(2,0)$ thus measures the load contrast between the central strip and the boundary strips,
and the mode $(0,2)$ measures the same along the other axis. In the great majority of states of both
classes the center is heavier. This is expected, because grains leave the system through the
boundary and the stationary density is lower there. The classes differ in the \emph{strength} of
this overload. The mean signed amplitude, normalized by $\sqrt{L^{2}q(1-q)}$ and averaged over both
orientations, is larger in absolute value before an extreme than in quiet states, by a factor of
1.85 at $L=32$ falling to 1.45 at $L=128$ (Appendix~\ref{app:pattern}).

\subsection{A single spatial precursor $R_2$}\label{sec:R2}

\noindent The two modes that stand out describe the same pattern in two orientations and map onto
each other under a rotation of the lattice by $90^{\circ}$. Following the rule of
Section~\ref{sec:modes}, we sum their normalized energies,
\begin{equation}\label{eq:R2}
R_2=R_{20}+R_{02}=\frac{a_{20}^{2}+a_{02}^{2}}{L^{2}q(1-q)} .
\end{equation}
$R_2$ is a single number per microstate, computed from a single frame and without training. It
measures the strength of the contrast between the center and the boundaries regardless of its sign
and orientation.

The quantity $R_2$ is separated from density algebraically but not statistically. Its AUC therefore does not by
itself show whether the variable contains information beyond the total number of grains. Conditioning
on the original number of grains provides this test (Section~\ref{sec:thinning}). The AUC of $R_2$
remains high within groups with the same $k$, reaching $0.92$-$0.97$, whereas density has an AUC of
exactly $1/2$ (Table~\ref{tbl5}). Thus the spatial signal persists without thinning and without a
network. The geometry of $R_2$ also accounts for the two scale scans in Section~\ref{sec:scale}.
Blocks of side $L/4$ coincide with the strips in which the mode of index 2 has a constant sign and
separate the central strip from the boundary strips. A block of side $L/2$ contains equal parts of the
two regions, so rearranging grains within it cancels the contrast (Appendix~\ref{app:pattern}).

\begin{table}[h!]
\begin{tabular*}{\linewidth}{@{\extracolsep{\fill}}llll@{}}
\toprule
$L$ & $\mathrm{AUC}(R_2)$ & $\mathrm{AUC}_{|k}(R_2)$ & $\mathrm{AUC}_{|k}(q)$ \\
\midrule
32  & 0.83 & 0.92 & 0.500 \\
64  & 0.85 & 0.95 & 0.500 \\
128 & 0.86 & 0.97 & 0.500 \\
\bottomrule
\end{tabular*}
\caption{The precursor $R_2$ with density removed by conditioning instead of thinning. The value of $\mathrm{AUC}_{|k}$ is computed based on Eq.~(\ref{eq:auck}) within groups of states with an identical number of grains
(Section~\ref{sec:thinning}).}\label{tbl5}
\end{table}

\subsection{Two macroscopic variables against the full microstate}\label{sec:two}

\noindent We next consider prediction from the raw images, where density is available. The logistic
model of Eq.~(\ref{eq:logit}) combines density with $R_2$. Its performance can be compared with the
network of Section~\ref{sec:predictors}, which is trained on the full microstate and therefore has
access to all $L^{2}$ sites (Table~\ref{tbl6}).

\begin{table}[h!]
\begin{tabular*}{\linewidth}{@{\extracolsep{\fill}}llllllll@{}}
\toprule
$L$ & $\mathrm{AUC}(q)$ & $\mathrm{AUC}(R_2)$ & $\mathrm{AUC}(q,R_2)$ &
$\Delta\mathrm{AUC}$ [95\% CI] & $\varepsilon(q)$ & $\varepsilon(q,R_2)$ &
$\varepsilon$ (network) \\
\midrule
32  & 0.954 & 0.826 & 0.985 & $+0.032$ [0.029, 0.034] & 0.228 & 0.118 & 0.097 \\
64  & 0.967 & 0.852 & 0.993 & $+0.026$ [0.023, 0.028] & 0.191 & 0.075 & 0.066 \\
128 & 0.974 & 0.856 & 0.996 & $+0.023$ [0.019, 0.026] & 0.170 & 0.059 & 0.061 \\
\bottomrule
\end{tabular*}
\caption{Two macroscopic variables against a convolutional network on the full microstate. The pair $(q,R_2)$ denotes the logistic model of Eq.~(\ref{eq:logit}). All
values are computed on the sets of Table~\ref{tbl1}, so $\mathrm{AUC}(q)$ and $\varepsilon(q)$ may
differ from the values in Table~\ref{tbl3}, which are computed on the full stream. Here the
negatives are a sample of $10\,000$ states and the threshold is chosen on the collected training set
(which is why $\varepsilon(q)$ at $L=64$ and 128 comes out slightly below the bound $\varepsilon^{*}$
determined on the stream). The confidence interval of
$\Delta\mathrm{AUC}=\mathrm{AUC}(q,R_2)-\mathrm{AUC}(q)$ comes from a paired bootstrap over test
examples. The difference is computed before rounding, so it may deviate by 0.001 from the difference
of the columns.}\label{tbl6}
\end{table}

\begin{figure}[h!]
  \centering
  \includegraphics[width=\linewidth]{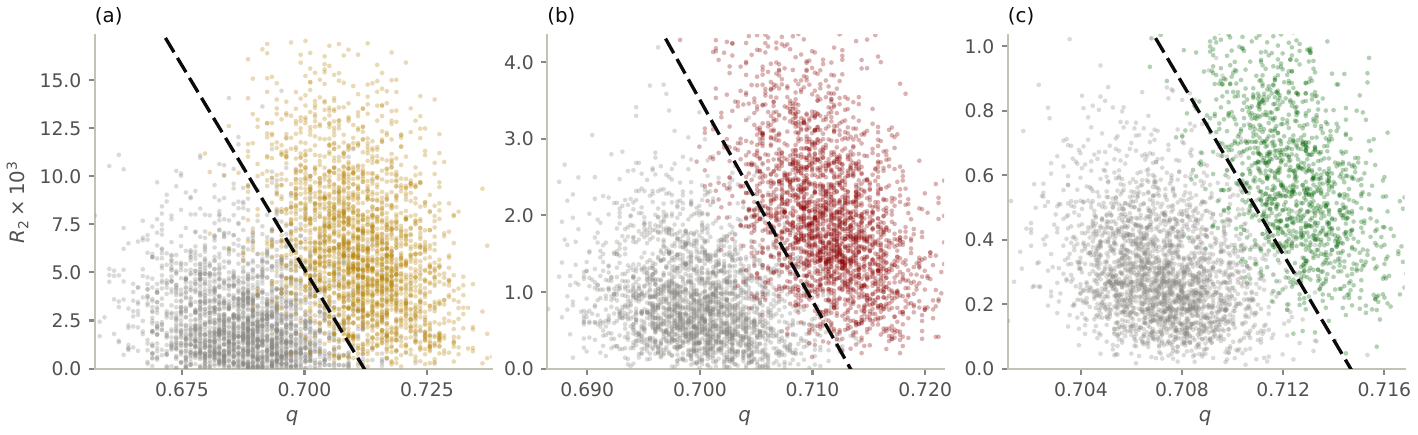}
  \caption{Test states in the plane of the two macroscopic variables, the density $q$ and the mode
  energy $R_2$, for (a) $L=32$, (b) $L=64$ and (c) $L=128$ (at most 2500 randomly chosen states per class). The
  values of $R_2$ are given in units of $10^{-3}$, and the range of the vertical axis differs
  between panels. Quiet states are gray, and states before an extreme have the color of their system
  size. The dashed line is the decision boundary of the logistic model, on which the model gives equal odds to both classes. Its coefficients were fitted
  on the training set and then frozen, so the line is shown on data the model has not seen. The alarm threshold used for $\varepsilon(q,R_2)$ in Table~\ref{tbl6} is a parallel line shifted toward lower load. The clouds are
  displaced both horizontally (higher load) and vertically (stronger contrast between the center and
  the boundaries), which is why two variables separate them better than either one
  alone.}\label{fig4}
\end{figure}

Adding $R_2$ to density lowers the error by about 0.11 at every size, which amounts to 84\% to 102\%
of the reduction achieved by the network on the full microstate (95\% confidence intervals from 80-88\%
at $L=32$ to 94-110\% at $L=128$). At $L=32$ and 64 the network remains better. At $L=128$ the two
variables match it within the spread between training seeds (0.059 against $0.061\pm0.008$).
Figure~\ref{fig4} shows why density alone is not enough. The clouds of the two classes are shifted
relative to each other along both axes, so the decision boundary is oblique.

The pair $(q,R_2)$ is not a complete description. On thinned images, where density is switched off,
$R_2$ alone reaches an AUC of 0.75-0.83, compared with 0.80-0.87 for the network. Neither of these
scores contains the other, however. Within groups of similar $R_2$ (20 quantiles) following the same Eq.~(\ref{eq:auck}), the network still
distinguishes the classes with an AUC of 0.70-0.75, so it also uses structure that $R_2$ does not
describe. Within groups of similar network score, $R_2$ still gives 0.54-0.64. The network is also
overfitted at the larger sizes (Appendix~\ref{app:net}), which in the comparison with the
two-variable model works in favor of that model.

\section{Discussion and conclusions}\label{sec:discussion}

\noindent The main result is that the global density $q$ does not completely describe the readiness of
the Manna model for an extreme avalanche. Density is a strong predictor, consistent with earlier
work~\cite{Shapoval2022}, but states with the same number of grains remain distinguishable. After thinning
to a fixed grain count, both a model-free statistic and a convolutional network perform above the
null level. Conditioning on the original number of grains gives the
same result and rules out a residual dependence on the thinning procedure. The remaining signal is
macroscopic: it appears on a scale proportional to the system size and is associated with a load
contrast between the center and the dissipative boundaries, represented by the degenerate cosine modes
$(2,0)$ and $(0,2)$. The corresponding scalar $R_2$ separates states with an identical grain count.
When density is restored, the pair $(q,R_2)$ recovers most of the error reduction obtained by the
network over density alone. Thus the number of grains is not a complete description of the system's
state with respect to extreme-avalanche prediction.

This result adds to the picture of SOC as an absorbing-state phase transition in which density is
the control parameter~\cite{Dickman1998,Dickman2000}. In a system with open boundaries the load is
not uniform, so besides its mean value it also matters how the load is distributed relative to the
boundaries. A natural interpretation, which we treat as a hypothesis and not as a measured result,
is that an avalanche developing in an overloaded interior has a longer way to the dissipative
boundary and can grow larger because of this. The hypothesis has some support. The variant of the
Manna model studied here is Abelian~\cite{Dhar1999}, and in Abelian sandpile models the balance of
grains in the stationary state relates the mean number of topplings to the Green's function of the
Laplacian with a dissipative boundary~\cite{Dhar1990}. This function is largest for grains added at
the center of the system. If the precursor is tied to the boundary, the fact that the maximum falls
at the same $\ell/L$ for all sizes cannot decide between the self-similarity of the critical state
and a load profile imposed by the finite system. The distance from the boundary scales with $L$ as
well. We therefore read this agreement as evidence against an artifact with a fixed size in lattice
sites. It does not prove self-similarity. Within the range studied, the role of spatial organization
does not weaken with system size. The conditional $\mathrm{AUC}_{|k}(R_2)$ increases, although the
relative difference in contrast between the classes decreases.

That extreme avalanches are preceded by spatial clustering has been reported
previously~\cite{ShapovalShnirman2004,ShapovalShnirman2006,Shapoval2022}. Here the spatial contribution is
isolated by construction, quantified against the null level, and localized in scale. The modal
decomposition then identifies the dominant pattern and reduces it to the scalar $R_2$. Comparing
with the network trained on the full microstate shows that this variable captures a large part of the information available
beyond density. The network is therefore useful mainly as a reference for the information content of
the microstate, while the modal analysis gives a direct physical description of the dominant spatial
component.

Several limitations remain. The pair $(q,R_2)$ does not exhaust the spatial information: on thinned
images the network still uses structure that $R_2$ does not describe. Whether density and this residual
structure can be replaced by a single quantity derived from the dynamics, as suggested by the boundary
hypothesis, remains to be established and is the subject of a separate paper. The present simulations
also cover only a fourfold range of system sizes, $L=32$-$128$. Because larger systems require more
computation and contain fewer extreme events, these data do not determine the large-$L$ limit or
separate a boundary effect from a property of the critical state. A model with bulk dissipation and
periodic boundaries, without a distinguished center, would provide a useful test. Finally, it remains
unclear whether the same spatial description applies to other sandpile models, including the
deterministic BTW model, which belongs to a different universality class~\cite{BenHur1996} and has
different finite-size predictability~\cite{Sapozhnikov2024}. Within the range studied here, however,
the predictive state depends on both the number of grains and their spatial arrangement, and most of
the network's additional information is captured by the single variable $R_2$.

\section*{Declaration of competing interest}

\noindent The authors declare that they have no known competing financial interests or personal
relationships that could have appeared to influence the work reported in this paper.

\section*{Data availability}
\noindent The code that supports the findings of this study is available from the corresponding
author upon request.

\section*{Acknowledgements}
J.S. is financially supported by the Foundation for Polish Science (FNP) grant ‘Centre for Credible AI’ No. FENG.02.01-IP.05-0058/24.

\section*{Declaration of generative AI and AI-assisted technologies in the manuscript preparation process}
During the preparation of this work the authors used Claude (Anthropic) and GPT (OpenAI) in order to assist with checking the reported numbers against the data and with language editing of
the manuscript. After using these tools, the authors reviewed and edited the content as needed and take full responsibility for the content of the published article.

\appendix

\renewcommand{\thefigure}{\Alph{section}.\arabic{figure}}
\renewcommand{\thetable}{\Alph{section}.\arabic{table}}
\renewcommand{\theequation}{\Alph{section}.\arabic{equation}}
\providecommand{\theHfigure}{}
\renewcommand{\theHfigure}{\Alph{section}.\arabic{figure}}
\providecommand{\theHtable}{}
\renewcommand{\theHtable}{\Alph{section}.\arabic{table}}
\providecommand{\theHequation}{}
\renewcommand{\theHequation}{\Alph{section}.\arabic{equation}}

\section{Scale scans: random level and the effect of thinning}\label{app:scans}
\setcounter{figure}{0}\setcounter{table}{0}\setcounter{equation}{0}

\noindent \textbf{Random level.} The quantity $E_{\text{rand}}$ of Eq.~(\ref{eq:delta}) is not
postulated but derived. Consider $K$ grains placed uniformly at random on $L^{2}$ sites, with at
most one grain per site. The number of grains $X_b$ in a block of $\ell^{2}$ sites then follows a
hypergeometric distribution with variance
\begin{equation}\label{eq:hyper}
\operatorname{Var}(X_b)=\ell^{2}\,q(1-q)\,\frac{L^{2}-\ell^{2}}{L^{2}-1},\qquad q=\frac{K}{L^{2}}.
\end{equation}
The density of a block is $\rho_b=X_b/\ell^{2}$ and the mean density of every block is $q$, so the
expected value of $V_\ell$ equals $\operatorname{Var}(X_b)/\ell^{4}$, which is the expression in
Eq.~(\ref{eq:delta}). For $\ell=1$ a block is a single site and $V_\ell=q(1-q)$ for every
configuration, so $\Delta=0$ by construction. At $L=32$ we therefore omit the point $\ell/L=1/32$.

\textbf{Effect of thinning.} Removing grains at random from an image with $k$ grains multiplies its
excess variance over the random level by $K(K-1)/[k(k-1)]$, equally at every scale $\ell$. This is
why we compare the shape of the curves $\Delta(\ell/L)$ between sizes. Positives, however, have more
grains than negatives, so their excess is weakened more, which adds a contribution to $\Delta$ that
has nothing to do with the arrangement. This is the same trace of the original number of grains that
is discussed in Section~\ref{sec:thinning}, and we remove it in the same way. We repeat the scan at
a fixed original number of grains. The difference
$\langle V_\ell\rangle_{+}-\langle V_\ell\rangle_{-}$ is computed within groups of identical $k$ and
averaged with weights $|P_k||N_k|/(|P_k|+|N_k|)$. In the repeated scan, normalized as in
Fig.~\ref{fig2}a, the maximum stays at $\ell/L=1/4$, and the relative excess at $\ell/L=1/2$ is
0.23-0.28 compared with 0.38-0.58 in the basic scan, with widely overlapping confidence intervals.

\section{Neural network: training, learning curves and generalization gap}\label{app:net}
\setcounter{figure}{0}\setcounter{table}{0}\setcounter{equation}{0}

\noindent \textbf{Architecture and training.} The head of the network of
Section~\ref{sec:predictors} computes, for each of the 32 channels of the last feature map, the mean
and the standard deviation over positions, which gives 64 numbers per image. The convolutions have
$3\times3$ kernels, and the blocks have between 16 and at most 64 channels. The last feature map has
size $4\times4$ at every $L$, and the network has between 14 and 79 thousand parameters. During
training we apply random elements of the symmetry group $D_4$ (rotations by multiples of
$90^{\circ}$ and reflections) as augmentation. The dynamics of the model is invariant under these
operations. We use the Adam optimizer~\cite{Kingma2015} (learning rate $10^{-3}$, weight decay
$10^{-4}$), a batch size of 64, at most 100 epochs with a patience of 10, and early stopping based
on the validation loss instead of the AUC. The loss function is the binary cross-entropy.

\begin{figure}[h!]
  \centering
  \includegraphics[width=\linewidth]{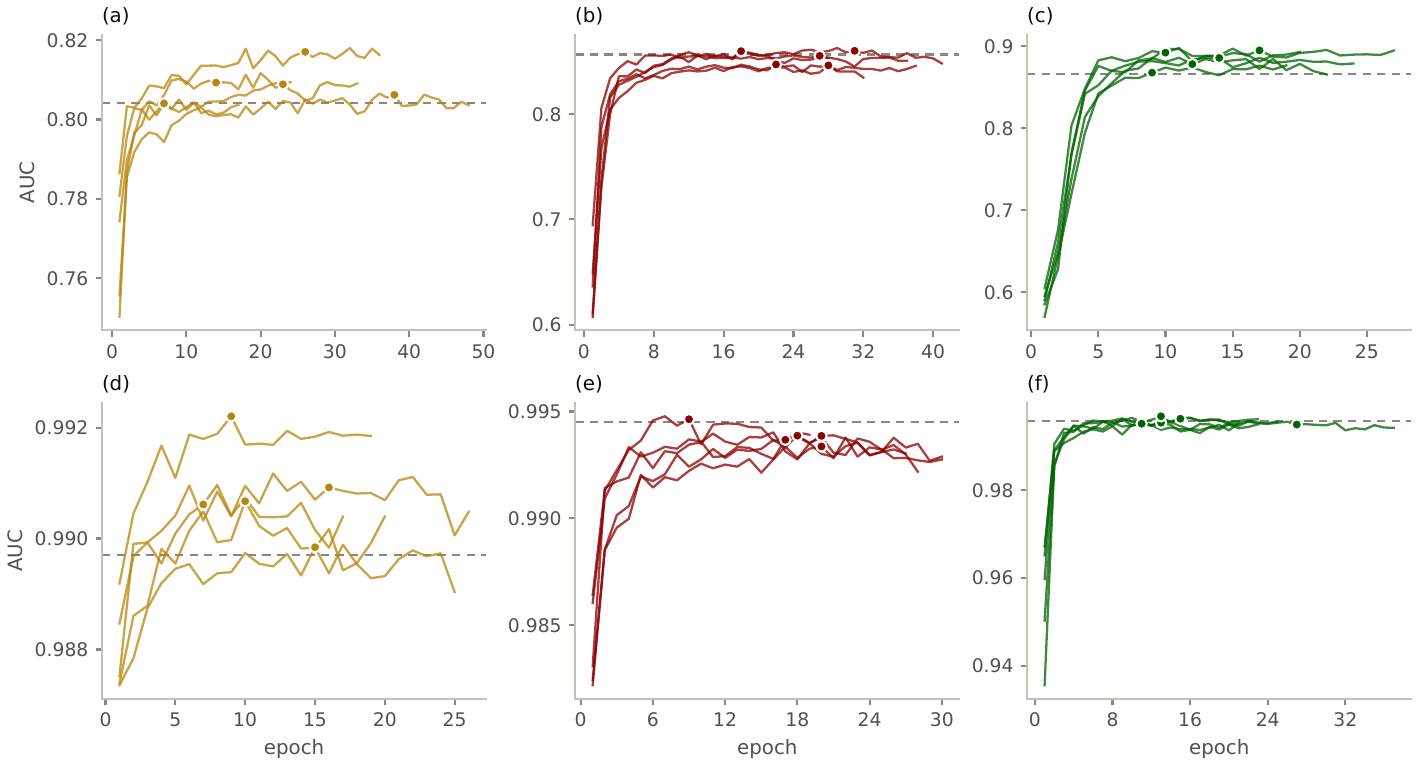}
  \caption{Learning curves of the convolutional network. Values of AUC on the validation part after successive
  training epochs (a)-(c) on thinned images and (d)-(f) on raw images, for $L=32$, 64 and 128 from left to right. Each line is one of the five seeds,
  and a dot marks the epoch from which the saved weights are taken. The dashed line is the AUC of
  these weights on the test set, averaged over seeds.}\label{figB1}
\end{figure}

\textbf{Learning curves.} Figure~\ref{figB1} shows the AUC on the validation part of the training
set after each epoch, separately for each of the five seeds. A dot marks the epoch with the smallest
validation loss, from which the weights used in the paper are taken.

None of the 45 training runs (thinned, shuffled and raw images, three system sizes, five seeds)
reached the limit of 100 epochs. The stopping criterion ended them after 11-48 epochs, on the flat
part of the curve. The mean AUC of the networks shown in the figure on the test set, which takes no
part in training, differs from the validation AUC by at most 0.018. The network trained on shuffled
images stays at the random level throughout training (validation AUC between 0.46 and 0.54).

\textbf{Generalization gap.} At $L=64$ and 128 the network is overfitted. On thinned images its AUC
on the examples it was trained on exceeds the test AUC by about 0.01 and 0.04 (the difference is negligible at
$L=32$), which lowers the result in Table~\ref{tbl4} and does not inflate it. On raw images its
error at $L=128$ rises from 0.034 on the examples it was trained on to 0.057 on its own validation
part, which contributes no gradient steps (in both cases with the threshold chosen on the same part). On the test set, with the threshold taken from the training set, it is 0.061 (Table~\ref{tbl6}). In the
comparison with the two-variable model this works in favor of that model.

\section{Spatial pattern: direction of the contrast and properties of $R_2$}\label{app:pattern}
\setcounter{figure}{0}\setcounter{table}{0}\setcounter{equation}{0}

\noindent \textbf{Direction of the contrast.} The sign of the normalized amplitude
$\hat a_{20}=a_{20}/\sqrt{L^{2}q(1-q)}$ tells which part of the system is heavier. A negative value
means a heavier center. In each of the two orientations separately, the amplitude is negative in at
least 87\% of quiet states and in at least 97\% of states before an extreme.

\textbf{$R_2$ and density.} Within each class the Spearman rank correlation between $R_2$ and $q$ is
negative, from $-0.14$ to $-0.25$ in quiet states and from $-0.36$ to $-0.48$ before an extreme.
Density thus partly masks the signal of $R_2$, and conditioning on the number of grains lifts this
mask. This explains the increase of $\mathrm{AUC}(R_2)$ in Table~\ref{tbl5}.

\textbf{$R_2$ and the scale of the signal.} After the grains are rearranged inside blocks of side
$L/4$, $R_2$ still distinguishes the classes (the AUC drops from $0.83$-$0.86$ to $0.73$-$0.76$).
After rearrangement inside blocks of side $L/2$ it no longer does ($0.49$-$0.51$), in line with the
geometric argument of Section~\ref{sec:R2}. The same argument explains the maximum of $\Delta$ at
$\ell/L=1/4$ and the largest increment of $F$ between $L/4$ and $L/2$.

\bibliographystyle{elsarticle-num}
\bibliography{ArXiv_SOC}
	
\end{document}